# Cognitive Analysis of Security Threats on Social Networking Services: Slovakia in need of stronger action

**Karol FABIÁN**
Department of Security Studies, Matej Bel University

**Jozef Michal MINTAL**
Department of Security Studies, Matej Bel University

**ABSTRACT**

This paper examines some of the ongoing research at the UMB Data and Society Lab hosted at the Faculty of Political Science and International Relations, Matej Bel University in Slovakia. It begins with an introduction on the necessity of security threat identification on social networking services (SNSs), done by states. The paper follows with a general overview of selected projects of the Lab in this field, and afterwards it introduces a use case study focused on the announcement of the UK snap general election 2017. The main aim of this paper is to demonstrate some of the possibilities of social networking services analysis in the field of international relations, with an emphasis on disinformation and the necessity of identifying novel digital actors in Slovakia. We also outline an easy custom system tasked to collect social media data, and afterwards process it using various cognitive analytic methods.

## 1. INTRODUCTION

Nowadays, modern, cheap and massively available information-communication technologies enabled a wide array of actors to participate and compete in the public space on equal terms with already established actors such as political leaders, government institutions, supranational organizations, etc. Thereby altering the interaction capacity [1] of these established actors not only in the "new digital realm", but also in the "psychical one". Nefarious actors are unfortunately not only present in the physical realm. In the recent past we were able to witness the active workings of various actors present on social networking services (SNSs), undermining the confidence of citizens in the state and democracy, by systematically spreading disinformation and half-truths. [2] [3]

Even if disinformation is not a new phenomenon, the intensity of the current disinformation wave is unlike any we have experienced in recent history, particularly because of the level and availability of modern technology. By utilizing SNS analytics and digital data points left by various nefarious actors, we are however able to map and better understand these actors, and thus create more meaningful mitigation strategies for the present and the future.

In the case of Slovakia, we can talk about far more vague approaches in detecting and countering these threats, sustained mainly thanks to the efforts of selected individuals and non-governmental organizations, such as Globsec, SSPI, Stratpol, blbec online, etc. [4]

Having the capacity to quickly and suitably analyze the behavior of individuals and groups on SNSs, while preserving user privacy and democratic conduct, and afterwards being able to mitigate the potentially malignant impact of some of this behavior on state security, is a necessary condition of ensuring internal security of any country in the 21st century, including that of Slovakia.

In supporting this endeavor, in 2018 selected members of the Secure Societies team at Matej Bel University (UMB) launched the UMB Data&Society Lab (UMB DSL). Housed at the Faculty of Political Science and International Relations, it is a multidisciplinary research lab engaged in research and education on the interaction of digital technology with society. The UMB DSL was launched thanks to the merger of two independently funded projects by IBM and the Scientific Grant Agency VEGA. Since its launch the Lab has established various partnerships with government, business, and civil society. Inter alia the Lab is a member of the Global Network of Internet & Society Research Centers – a group of academic institutions with a focus on interdisciplinary research on the development, social impact, policy implications, and legal issues concerning the Internet. In its research the Lab is focusing on three main areas: Digital Social Influence, Digital Security Threats, and Implications of A.I. for society, in these domains it aims to broaden the theoretical and empirical foundation of internet and society research, and to contribute to a better understanding of the digital society. Current Research Projects of the Lab include Cognitive Analytics for Real-World Security Threats (CARWST), supported by the IBM Global University programs; The Possibilities of the Cognitive Analysis of the Security Threats on the Social Networks in the Cyber-Space, supported by the Scientific Grant Agency VEGA; Data & Society Lab Endeavour New Way Of Utilizing AI 2020, supported by IBM CE.

## 2. USE CASE STUDY

In this case study we used a multi-method approach in researching digital attention during a selected discussion on the social networking service Twitter, with the discussion in question being the announcement of the 2017 snap general election in the UK. For the collection of the discussion network, we used a topic-based sampling method, with hashtag identification being manually selected based on top trending topic related hashtags, utilizing Trendsmap Analytics (Trendsmap Pty Ltd, Australia). Manual relatedness identification was done by two independent coders, with the most suitable hashtag being identified to be #GE2017. Data collection was performed on April 21, 2017 using the Twitter API. The collected dataset included (n)=12.434 users and (m)=19.888 connections among the users, with connections representing replies to, mentions and follows. The dataset was then exported into Gephi (Gephi Consortium, France).

Using the obtained dataset, we first calculated the network modularity using the Louvain method [5], with the modularity being =0,742 and (n)=1.250 detected community structures, this indicates a higher level of separation of communities and thus a lower interconnectivity of the whole network. For influential actor identification we calculated the eigenvector centrality metric of all nodes in the network. On the collected tweets we conducted a semantic analysis using the Opinion Lexion by Hu and Liu [6] calculating the number of word- mentions as well as their pertaining salience value. The whole network was then visualized using ForceAtlas2.

**Table 1.** *Top twitter accounts by eigenvector*

| Twitter handle | determined owner | Eigenvector |
|---|---|---|
| @jeremycorbyn | Jeremy Corbyn | 0,015 |
| @peoplesmomentum | Momentum | 0,010 |
| @owenjones84 | Owen Jones | 0,008 |
| @labourlewis | Clive Lewis | 0,004 |
| *retracted* | *retracted* | 0,003 |
| *retracted* | *retracted* | 0,003 |
| @uklabour | *retracted* | *0,002* |
| *retracted* | *retracted* | *0,002* |
| *retracted* | *retracted* | *0,002* |
| *retracted* | *retracted* | *0,002* |
| @richardburgon | Richard Burgon | 0,002 |
| @barrygardiner | Barry Gardiner | 0,002 |
| @theresa_may | Theresa May | 0,002 |

Despite the then dominance of the Conservative party in the polls, the SNS primacy, in this researched Twitter discussion, belonged to the Labour party, with the official twitter account of Jeremy Corbyn having the highest eigenvector value, the account of Labour MP Owen Jones placing third, and the official account of the Labour party placing sixth. This is in stark contrast to the highest placing officially associated conservative account being only 13[th], with the account belonging to Theresa May. (Table 1.)

Results from the semantic analysis were in the same line with the eigenvector results, the Labour party was the more talk about party with the word Labour being mentioned a total of 1.762 times in the tweets and having a salience value of 0,008 compared with the word Tory being mentioned 1.430 times and having a salience value of 0,007. (Table 2.) However, it would be erroneous to automatically transpose this digital "popularity" into real world election ballot dominance. First the mentioned metrics such as eigenvector value and salience are context neutral, second SNSs are in no way's representative of the general population, with social networking services usage having a lower penetration among some demographic groups, also other factors are further skewing the image, inter alia hyper-active users generating an unproportionally large number of content [7].

However, in the context of this paper it is interesting to also look at novel digital actors present in the dataset. In this instance for example an account belonging to a mother of two from Manchester, with no proven direct political links or celebrity status had a larger eigenvector value (=0,003) than the official Twitter account of then prime minister Theresa May. (Table 1. personal twitter handles retracted due to privacy regulation).

**Table 2.** *Top word count and word salience value*

| Word | number of mentions | Salience value |
|---|---|---|
| Rt | 15.479 | 0,008 |
| Ge2017 | 14.131 | 0,008 |
| Vote | 2.313 | 0,010 |
| Brexit | 2.310 | 0,009 |
| Amp | 1.933 | 0,009 |
| Labour | 1.762 | 0,008 |
| *Retracted* | 1.748 | 0,008 |
| Voting | 1.595 | 0,008 |
| Tory | 1.430 | 0,007 |

Would this account have been belonged to a nefarious foreign/domestic actor it might have constituted a potential security threat, for example in the form of foreign election interference. Thus, it is vital for the state to have the capacity to quickly but also suitably, taking into account inter alia user privacy and adhering to a democratic conduct, analyze the "new digital realm". Unfortunately, this as we have already mentioned, is not the case in Slovakia where government agencies tasked with monitoring the digital realm, such as the Strategic Communication Unit at the Ministry of Foreign and European Affairs of the Slovak Republic, are dramatically understaffed and lack the basic necessary technical infrastructure for such SNSs monitoring. In recent years, another hurdle in conducting meaningful analysis has also crystalized, that is the lack of data access provided by SNS platforms such as Facebook and Twitter. This is on one hand understandable, due to the potential of misuses; however, the current situation might be implicitly privileging larger countries which have the necessary technical and personal capacity to obtain the data in other ways such as illegal data scraping, exclusive data access contracts, buying expensive dataset, etc. and thus avoid some of the data restrictions officially imposed by the platforms. [8]

### 3. PROPOSING INTEGRATIVE APPROACHES

As previously mentioned, Slovak government agencies tasked with monitoring SNSs and conducting strategic communication lack the necessary technical infrastructure to effectively deliver meaningful monitoring, we thus propose an easy setup hosted at the Crisis Management Center (CEKR) at the Faculty of Political Science and International Relations UMB, tasked with collecting social media data, and afterwards processing it using various cognitive analytics methods. The system utilizes a CEKR web server running Cogniware Data Collector (CWDC) [9], a state-of-art data collector by Cogniware (Cogniware s.r.o., Czechia) which enables the collection of various publicly available data sources form social networking services APIs, RSS channels, and is also able to amend these with various other online and offline data. Output records are either in JSON or XML. CWDC is in our setup connect to the IBM Watson Explorer Content Analytics (IBM, USA) to which CWDC exports the data, this component houses some of the cognitive analytic features such as sentiment analysis, and deviation notification. Additional pipelines can export CWDC data as stand-alone files, for manual analysis in Gephi (Gephi Consortium, France) and other network analysis and visualization software, as well as to IBM SPSS Modeler (IBM, USA) for creating inter alia predictive analytic models that can be utilized for example to detect bots on social networking services. As the whole setup is housed at the Faculty of Political Science and International Relations UMB, it can also tap into the policy experience of selected faculty members, and thus also provide meaningful policy recommendations in easy to understand reports. There are however various limitations in fully adopting this strategy which need to be first overcome in order to thoroughly adopt this proposed approach. The limitations can be categorized as internal mainly due software license issues as well as external mainly due to SNSs API restrictions.

## 4. CONCLUSIONS

The changes of interaction capacities of established institutions, such as that of states, political leaders, etc. based on the democratization of the public space of the digital realm, which was created by modern information communication technology in general and social networking services in particular, has not only had positive effects on society and security but also, demonstratively, negative ones, with these having been of particular focus in the recent years.

Thus having the capacity to quickly but suitably analyze the behavior of individuals and groups on SNSs, while at the same time preserving user privacy and democratic conduct, and having the capacity to mitigate the potentially malignant impact of some of these behaviors on state security is a necessary condition of ensuring internal security of any country in the 21st century. In this paper we sketched out this necessity with a focus on novel digital actor identification, utilizing a use case in the form of a Twitter discussion related to the announcement of the UK snap election 2017. We demonstrated that novel digital actors, might wield great online popularity, and thus potentially pose great risks to state security if weaponized. In this paper we also roughly outlined the current underdeveloped and underfunded situation in Slovakia in the social media data and analysis domain and introduced potential solutions that are being work on in our Lab, with the most promising being a custom crawler and cognitive analytics setup.

## 5. DECLARATION

This paper was supported by the Scientific Grant Agency VEGA project n. 1/0433/18, and by an IBM Global University Grant - Data & Society Lab Endeavour New Way of Utilizing AI 2020, IBM CEE. The funders had no role in study design, data collection and analysis, decision to publish, or preparation of the manuscript. Parts of the research were published as a master thesis of the second author – J.M. Mintal, thesis title "Jednotlivec a jeho schopnosť ovplyvňovať aktérov sociálnej siete". The data are available upon request from the authors.